\documentclass[a4paper,12pt]{article}
\usepackage{bm}
\usepackage{graphicx}

\title{Intrinsic spin currents in bulk 
       noncentrosymmetric ferromagnets}

\author{I. Turek \\
\textit{Institute of Physics of Materials, 
        Czech Academy of Sciences,} \\
\textit{Zizkova 22, CZ-61600 Brno, Czech Republic} \\
\texttt{turek@ipm.cz} }

\date{\today}

\begin{document}

\maketitle

\begin{abstract}
Intrinsic spin currents are encountered in noncentrosymmetric
crystals without any external electric fields; these currents
are caused by spin-orbit interaction.
In this paper, various theoretical aspects of this phenomenon in
bulk ferromagnets are studied by using group theory, perturbation
expansion, and calculations for model and real systems.
The group-theoretical analysis of the spin-current tensor shows
that the absence of space-inversion symmetry is not a sufficient
condition for appearance of the intrinsic spin currents.
The perturbation expansion proves that in the regime of exchange
splitting dominating over spin-orbit interaction, the spin
polarization of the intrinsic currents is nearly perpendicular to
the direction of magnetization.
First-principles calculations are carried out for NiMnSb and
CoMnFeSi ferromagnetic compounds, both featured by a tetrahedral
crystallographic point group.
The dependence of the spin-current tensor on the direction of
magnetization is approximated by a simple quadratic formula
containing two constants; the relative error of this
approximation is found as small as a few percent for both
compounds.
\end{abstract}

\newpage

\section{Introduction\label{s_intr}}

Spin currents generated in various materials and devices by
external electric fields belong to the most important topics
in spintronics \cite{r_2012_tz, r_2012_mvs}.
In this context, the well-known phenomenon is the spin Hall
effect in nonmagnetic systems \cite{r_2015_svw}, where the
spin currents transversal to the electric field are encountered.
Moreover, spin-polarized longitudinal currents can be found in
certain nonmagnetic solids \cite{r_2015_wsc}.
Similarly, spin polarization of electron currents in magnetically
ordered systems attracts ongoing interest as well; this involves
both systems with traditional magnetic orders \cite{r_2014_zck,
r_2015_skw, r_2017_zsy, r_2019_als, r_2023_syh, r_2024_kdb}
and the recently introduced altermagnets \cite{r_2021_hsv,
r_2021_szl, r_2022_shg, r_2022_ssj, r_2022_it}. 

Besides the spin currents induced by applied electric fields, it
was predicted two decades ago that systems lacking space-inversion
symmetry (noncentrosymmetric systems) can exhibit nonzero spin
currents even without an external perturbation \cite{r_2003_eir}.
These intrinsic spin currents arise due to spin-orbit (SO)
interaction; their consequences for properties of nonmagnetic
semiconductors \cite{r_2004_eir} and for improved definitions of
the spin-current operator \cite{r_2006_wxs, r_2006_szx} were
discussed in the literature.

The intrinsic spin currents are also relevant for
noncentrosymmetric ferromagnets, especially due to their close
connection to the Dzyaloshinskii-Moriya (DM) interaction
\cite{r_2016_kka, r_2017_fbm}.
This interaction can lead to an instability of the ferromagnetic
order resulting in a formation of noncollinear magnetic structures,
such as magnetic skyrmions \cite{r_2016_sm, r_2017_jhh}. 
Full details of the relation between the spin currents and the
DM interaction remain yet to be clarified; nevertheless, the
rough equivalence of both quantities seems to be confirmed by
existing \emph{ab initio} calculations using different approaches
\cite{r_2016_kka, r_2017_fbm, r_2014_fbm_c, r_2017_me, r_2019_mpe}.

The present paper addresses several topics in the theory of
the intrinsic spin currents in bulk noncentrosymmetric
ferromagnets.
The focus is on systems with the exchange splitting dominating
over the strength of SO interaction. 
Different tools are used in the study: perturbation expansion,
group theory, and computations on model as well as \emph{ab initio}
levels.
Two particular Heusler-like systems are chosen for
first-principles calculations,
namely, NiMnSb with $C1_b$ structure \cite{r_2002_gdp_h} and
CoMnFeSi with LiMgPdSn-type structure \cite{r_2015_bmr}, both
featured by the crystallographic point group $\bar{4}3m$.
Special attention is paid to the dependence of the spin currents
on direction of magnetization.
The main aim is to get insight into the properties of intrinsic
spin currents for a comparison with the known features of 
field-induced spin currents \cite{r_2019_als, r_2023_syh,
r_2024_kdb} and of the DM interaction \cite{r_2017_fbm}.

%\newpage

\section{Methods\label{s_meth}}

We assume a bulk ferromagnet with perfect translational invariance,
described by an effective one-electron Hamiltonian
\begin{equation}
H = \bar{H} + H^\mathrm{ex} + H^\mathrm{SO} ,
\label{eq_hdef}
\end{equation}
where the first term $\bar{H}$ denotes a spin-independent part
containing both local (site-diagonal) and nonlocal (hopping)
contributions, the second term $H^\mathrm{ex}$ denotes a
spin-dependent local exchange splitting part, and the third term
$H^\mathrm{SO}$ denotes a spin-dependent local SO interaction.
The exchange splitting points along a fixed unit vector
$\mathbf{m} = ( m_x , m_y , m_z )$, which can be identified with
the direction of magnetization of the ferromagnet.

The spin current is a tensor of rank two, $Q_{\kappa \lambda}$
($\kappa , \lambda \in \{ x, y, z \}$), where the first subscript
$\kappa$ refers to the spin polarization while the second subscript
$\lambda$ refers to the spatial direction of the current.
Its value at zero temperature is given by
\begin{equation}
Q_{\kappa \lambda} = \Omega^{-1} \sum_j
\langle j | \sigma_\kappa V_\lambda | j \rangle ,
\label{eq_qdef}
\end{equation}
where $\Omega$ denotes the volume of a large finite crystal
with periodic boundary conditions and the symbol $j$ labels
all eigenstates with eigenvectors $| j \rangle$ (normalized to
unity in the crystal volume $\Omega$, $\langle j | j \rangle = 1$)
and eigenvalues $E_j$ of the Hamiltonian $H$.
The summation extends only over the occupied eigenstates with
energies $E_j$ not exceeding the Fermi energy $E_\mathrm{F}$
of the system.
The symbol $\sigma_\kappa$ in Eq.~(\ref{eq_qdef}) denotes the Pauli
spin matrix and the symbol $V_\lambda$ denotes the velocity
(current) operator, that is spin-independent, derived from the
hopping part of the term $\bar{H}$.

Besides the complete tensor $Q_{\kappa \lambda}$, we have also
studied its projection on the unit vector $\mathbf{m}$.
This leads to a projection vector with components $P_\lambda$
defined by
\begin{equation}
P_\lambda = \sum_\kappa m_\kappa Q_{\kappa \lambda} ,
\label{eq_pdef}
\end{equation}
which bears information about mutual orientation of the spin
polarization of the intrinsic current and the magnetization
direction.

The previous expression for the spin current, Eq.~(\ref{eq_qdef}),
can be reformulated as a complex integral
\begin{equation}
Q_{\kappa \lambda} = \frac{1}{2 \pi i \Omega} \int_C \mathrm{Tr}
\{ \sigma_\kappa V_\lambda G(\epsilon) \} d\epsilon ,
\label{eq_qcint}
\end{equation}
where the trace (Tr) refers to the Hilbert space of the whole
finite crystal, $\epsilon$ denotes a complex energy variable,
the symbol $G(\epsilon) = (\epsilon - H)^{-1}$ is the resolvent
of the Hamiltonian $H$, and the integration path $C$ starts and
ends at the Fermi energy $E_\mathrm{F}$ and it contains the
occupied part of the spectrum of the Hamiltonian $H$ in its
interior.
The latter expression is used in an analysis of perturbation
expansion, see Section~\ref{ss_psc}.

The vanishing of the intrinsic spin currents for centrosymmetric
systems reflects the fact that the spin-current operator
$\sigma_\kappa V_\lambda$ changes its sign under space inversion.
However, detailed information about the shape of the spin-current
tensor $Q_{\kappa \lambda}$ for arbitrary nonmagnetic and magnetic
crystals demands the use of group theory.
This was worked out to many details for various tensor quantities
in the past \cite{r_2015_skw, r_1963_rrb, r_1966_whk, r_2019_gee};
in this study we applied an approach based on projection
superoperators \cite{r_2022_it}.

The first-principles evaluation of the spin-current tensor was
carried out using the fully relativistic tight-binding linear
muffin-tin orbital (TB-LMTO) method \cite{r_2014_tkd, r_2019_tkd}.
This requires a modification of Eq.~(\ref{eq_qcint}) into the form
\begin{equation}
Q_{\kappa \lambda} = \frac{1}{2 \pi i \Omega} \int_C \mathrm{Tr}
\{ \sigma_\kappa v_\lambda g(\epsilon) \} d\epsilon ,
\label{eq_qlmto}
\end{equation}
where $v_\lambda$ and $g(\epsilon)$ denote respectively the
effective velocity operator and the auxiliary resolvent of the
TB-LMTO technique.
It can be proved that Eq.~(\ref{eq_qlmto}) is invariant with
respect to the screening transformations of the TB-LMTO method.
Numerical implementation and computational details used were
similar to those described in the previous studies of the anomalous
\cite{r_2014_tkd} and spin \cite{r_2019_tkd} Hall conductivities.

%\newpage

\section{Results and discussion\label{s_redi}}

\subsection{Polarization of spin currents\label{ss_psc}}

Let us start with an analysis of perturbation expansion of the
spin-current tensor corresponding to inclusion of the term
$H^\mathrm{SO}$ into the Hamiltonian $H = H_0 + H^\mathrm{SO}$
with the reference nonrelativistic Hamiltonian $H_0 = \bar{H}
+ H^\mathrm{ex}$.
The resolvent $G(\epsilon)$ and the reference resolvent
$G_0(\epsilon) = (\epsilon - H_0)^{-1}$ are related by
\begin{equation}
G(\epsilon) = G_0(\epsilon) + G_0(\epsilon) H^\mathrm{SO}
 G_0(\epsilon) + \dots .
\label{eq_dyson}
\end{equation}
Substitution of this relation into Eq.~(\ref{eq_qcint}) and using
the properties of nonrelativistic systems results in the
well-known linear scaling of the spin-current tensor components
$Q_{\kappa \lambda}$ with the strength of SO interaction, as has
also been found in recent calculations for realistic systems
\cite{r_2017_fbm}.

However, the scaling of the projection vector $P_\lambda$,
Eq.~(\ref{eq_pdef}), is featured by an exponent higher than unity.
This can be seen from the expansion
$P_\lambda = P^{(0)}_\lambda + P^{(1)}_\lambda + \dots$, where
the individual terms correspond to those in Eq.~(\ref{eq_dyson}).
We get not only $P^{(0)}_\lambda = 0$ (needed for a linear
scaling), but also $P^{(1)}_\lambda = 0$ (leading to a quadratic
scaling).
The latter follows from Eqs.~(\ref{eq_pdef}), (\ref{eq_qcint}),
and (\ref{eq_dyson}), which yield 
\begin{equation}
P^{(1)}_\lambda = \frac{1}{2 \pi i \Omega}
 \int_C F(\epsilon) d\epsilon ,
\quad
F(\epsilon) = \mathrm{Tr}
\{ ( \mathbf{m} \cdot \bm{\sigma} ) V_\lambda G_0(\epsilon)
 H^\mathrm{SO} G_0(\epsilon) \} ,
\label{eq_p1}
\end{equation}
where we abbreviated $\bm{\sigma} =
( \sigma_x , \sigma_y , \sigma_z )$
and $\mathbf{m} \cdot \bm{\sigma} = 
\sum_\kappa m_\kappa \sigma_\kappa$.
Subsequently, one obtains
\begin{equation}
F(\epsilon) =
 \mathrm{Tr} \{ G_0(\epsilon) V_\lambda G_0(\epsilon)
 H^\mathrm{SO} ( \mathbf{m} \cdot \bm{\sigma} ) \} = 0 .
\label{eq_tr0}
\end{equation}
In the first step, we used the cyclic invariance of the trace
together with the fact that the operator
$\mathbf{m} \cdot \bm{\sigma}$ commutes with the reference
Hamiltonian $H_0$ and with its resolvent $G_0(\epsilon)$.
In the second step, we used the facts that the operator
$H^\mathrm{SO} ( \mathbf{m} \cdot \bm{\sigma} )$ is 
site-diagonal and that the on-site blocks of the operator
$G_0(\epsilon) V_\lambda G_0(\epsilon)$ vanish, see Eq.~(23)
in Ref.~\cite{r_2014_tkd}. 
This proves Eq.~(\ref{eq_tr0}) and, consequently, the vanishing
of $P^{(1)}_\lambda$.

\begin{figure}[h]
\begin{center}
\includegraphics[width=0.5\textwidth]{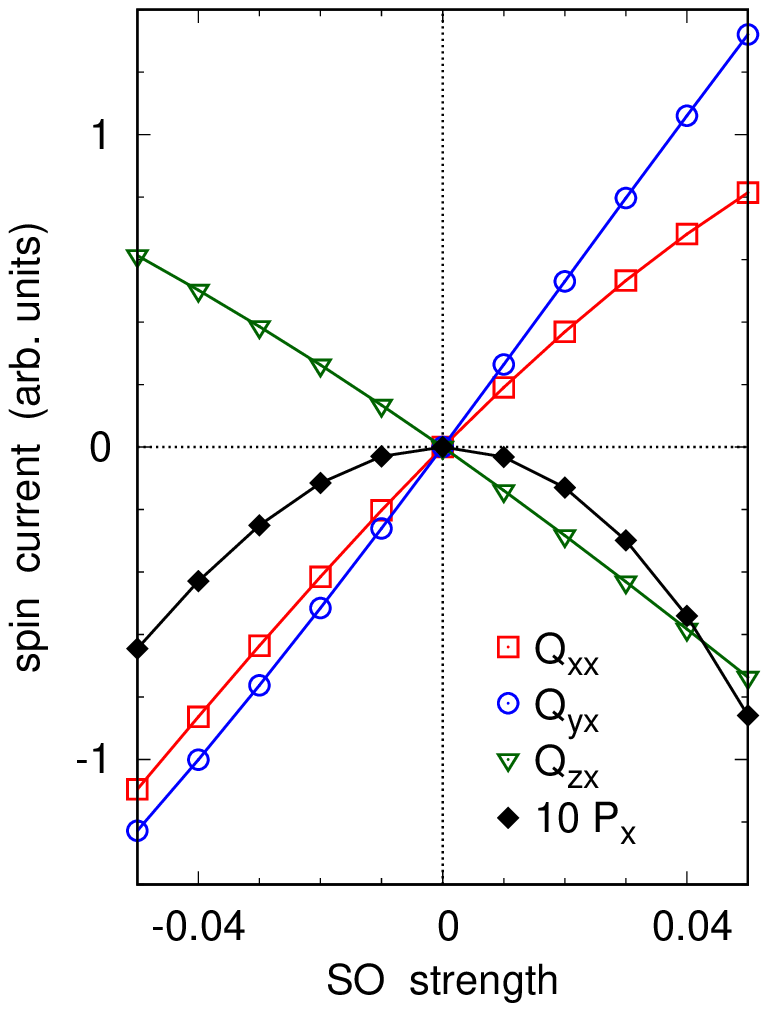}
\caption{Spin currents in the one-dimensional model as functions
of the strength of SO interaction: components of the
spin-current tensor $Q_{\kappa x}$ (open symbols) and of the
projection vector $P_x$ (full diamonds, magnified by a factor
of ten).
\label{fig_scaling}}
\end{center}
\end{figure}

This conclusion is supported by calculations for a
tight-binding model of a one-dimensional chain along the $x$
axis, described in detail in Appendix~\ref{app_1d}.
In this model, the values of index $\lambda$ are confined to
$\lambda = x$ and the remaining components of the spin-current
tensor and of the projection vector are displayed in
Fig.~\ref{fig_scaling} as functions of the strength of SO
interaction.
One can see clearly the linear scaling of the tensor components
$Q_{xx}$, $Q_{yx}$, and $Q_{zx}$, in contrast to the quadratic
scaling of the only component $P_x$ of the projection vector.

This result means that in the regime of exchange splitting
dominating over SO interaction, the spin polarization of
the intrinsic spin currents is practically perpendicular to
the magnetization direction.
This feature contrasts the properties of field-induced spin
currents in collinear nonrelativistic magnets, where the
spin polarization of the current is always strictly parallel
to the direction of magnetic moments, both in ferromagnets
and altermagnets \cite{r_2024_kdb, r_2022_it}.
In ferromagnets with nonzero SO interaction, the spin
polarization of the transversal currents (spin Hall effect)
also contains a nonnegligible component parallel to the
magnetization direction \cite{r_2019_als, r_2023_syh}.

\subsection{Group-theoretical analysis\label{ss_gta}}

For crystals without spontaneous magnetic order, the shape of
the spin-current tensor $Q_{\kappa \lambda}$ is dictated by
the crystallographic point group.
This tensor vanishes for all centrosymmetric point groups;
for all noncentrosymmetric point groups, the numbers of
independent components of the tensor are listed in
Table~\ref{tab_ncs_nm}.
One can see that symmetry reduction of the system leads in
general to an increase of the number of independent components,
as expected.
However, one also finds that three particular groups ($\bar{6}$,
$\bar{6}m2$,
$\bar{4}3m$) yield vanishing spin-current tensors, despite the
absence of space inversion.
Note that these three groups form a special class from the
viewpoint of chiral structures and related physical properties
\cite{r_2022_fkf}.

\begin{table}
\caption{%
Numbers of independent components of the spin-current tensor
for all noncentrosymmetric crystallographic point groups of
nonmagnetic solids.
\label{tab_ncs_nm}}
\begin{center}
\begin{tabular}{lc}
\hline
point group & no.\ of components \\
\hline
$1$ & 9 \\
$2$ & 5 \\
$m$ & 4 \\
$222$, $3$, $4$, $6$ & 3 \\
$2mm$, $32$, $422$, $\bar{4}$, $622$ & 2 \\
$3m$, $4mm$, $\bar{4}2m$, $6mm$, $23$, $432$ & 1 \\
$\bar{6}$, $\bar{6}m2$, $\bar{4}3m$ & 0 \\
\hline
\end{tabular}
\end{center}
\end{table}

\begin{table}
\caption{%
Numbers of independent components of the spin-current tensor
for all noncentrosymmetric magnetic point groups compatible
with ferromagnetism.
\label{tab_ncs_fm}}
\begin{center}
\begin{tabular}{lc}
\hline
magnetic point group & no.\ of components \\
\hline
$1$ & 9 \\
$2$, $2'$ & 5 \\
$m$, $m'$ & 4 \\
$3$, $4$, $6$, $2'2'2$ & 3 \\
$\bar{4}$, $m'm'2$, $m'm2'$, $32'$, $42'2'$, $62'2'$ & 2 \\
$3m'$, $4m'm'$, $\bar{4}2'm'$, $6m'm'$ & 1 \\
$\bar{6}$, $\bar{6}m'2'$ & 0 \\
\hline
\end{tabular}
\end{center}
\end{table}

For magnetically ordered crystals, one has to use the magnetic
point groups.
Among the total number of 122 magnetic point groups, only 21
groups are noncentrosymmetric and compatible with
ferromagnetic order.
For all these groups, the numbers of independent components of
the spin-current tensor are listed in Table~\ref{tab_ncs_fm}.
One can see that most of these groups support the existence of
nonzero spin currents; the only exceptions are two hexagonal
groups ($\bar{6}$, $\bar{6}m'2'$).
Note that both groups also lead to vanishing SO torques
generated by external electric fields \cite{r_2016_wcs}.

One can thus observe that both in nonmagnetic and ferromagnetic
crystals, missing space-inversion symmetry does not represent
a sufficient condition for the presence of nonzero intrinsic spin
currents.

\subsection{Spin currents in NiMnSb and CoMnFeSi
            alloys\label{ss_scha}}

The real systems selected for \emph{ab initio} study are NiMnSb
with $C1_b$ structure \cite{r_2002_gdp_h} and CoMnFeSi with
LiMgPdSn-type structure \cite{r_2015_bmr}, both belonging to the
crystallographic space group $F\bar{4}3m$ (No.\ 216).
Their structures (related closely to the standard Heusler
structure) are derived from an fcc Bravais lattice with four
sublattices, labelled $A$, $B$, $C$, and $D$, which are shifted
mutually along the (111) direction of the cubic lattice.
The basis vectors of these sublattices are: $A (0,0,0)$,
$B (0.25, 0.25, 0.25)$, $C (0.5, 0.5, 0.5)$, and
$D (0.75, 0.75, 0.75)$, all in units of the fcc lattice parameter. 
The sublattice occupation in both compounds is:
Ni($A$)Mn($B$)Sb($D$) (with $C$ sublattice empty) and
Co($A$)Mn($B$)Fe($C$)Si($D$).
Their total magnetic moment is around $4~\mu_\mathrm{B}$ per
formula unit \cite{r_2002_gdp_h, r_2015_bmr}, indicating that both
systems are in a regime of strong exchange splitting.

\begin{table}
\caption{%
Independent components $Q_{\kappa \lambda}$ of the spin-current
tensor in NiMnSb and CoMnFeSi alloys for magnetization parallel
to three high-symmetry directions.
The values obtained from the least-squares fit are displayed in
parenthesis.
\label{tab_qcalc}}
\begin{center}
\begin{tabular}{lccc}
\hline
magnetization & indices & $Q_{\kappa \lambda}$ (meV/nm$^2$)
 & $Q_{\kappa \lambda}$ (meV/nm$^2$) \\
direction & $\kappa \lambda$ & NiMnSb & CoMnFeSi \\
\hline
(001) & $yy$ & $-0.546 (-0.537)$ & $1.202 (1.210)$ \\
(111) & $xy$ & $-0.179 (-0.185)$ & $0.409 (0.404)$ \\
(101) & $zz$ & $-0.261 (-0.268)$ & $0.615 (0.605)$ \\
(101) & $zx$ & $-0.280 (-0.277)$ & $0.600 (0.606)$ \\
\hline
\end{tabular}
\end{center}
\end{table}

The crystallographic point group of these systems (with
ferromagnetic order ignored) is $\bar{4}3m$, which belongs to
the three special noncentrosymmetric groups leading to vanishing
spin currents, see Table~\ref{tab_ncs_nm}.
This means that nonzero spin currents appearing in the
ferromagnetic state have to be ascribed solely to the magnetic
order.
The calculated spin-current tensor for the magnetization direction
$\mathbf{m}$ along three high-symmetry directions of the lattice
is shown in Table~\ref{tab_qcalc}.
For $\mathbf{m} || (001)$, the magnetic point group is
$\bar{4}2'm'$ and the only independent nonzero tensor
component is $Q_{xx} = -Q_{yy}$.
For $\mathbf{m} || (111)$, the group is $3m'$ and the only
component is $Q_{xy} = Q_{yz} = Q_{zx} = -Q_{yx} = -Q_{zy} =
-Q_{xz}$.
For $\mathbf{m} || (101)$, the group is $m'm2'$ leading to two
slightly different components, namely, $Q_{xx} = -Q_{zz}$ and
$Q_{xz} = -Q_{zx}$.

\begin{figure}[h]
\begin{center}
\includegraphics[width=0.6\textwidth]{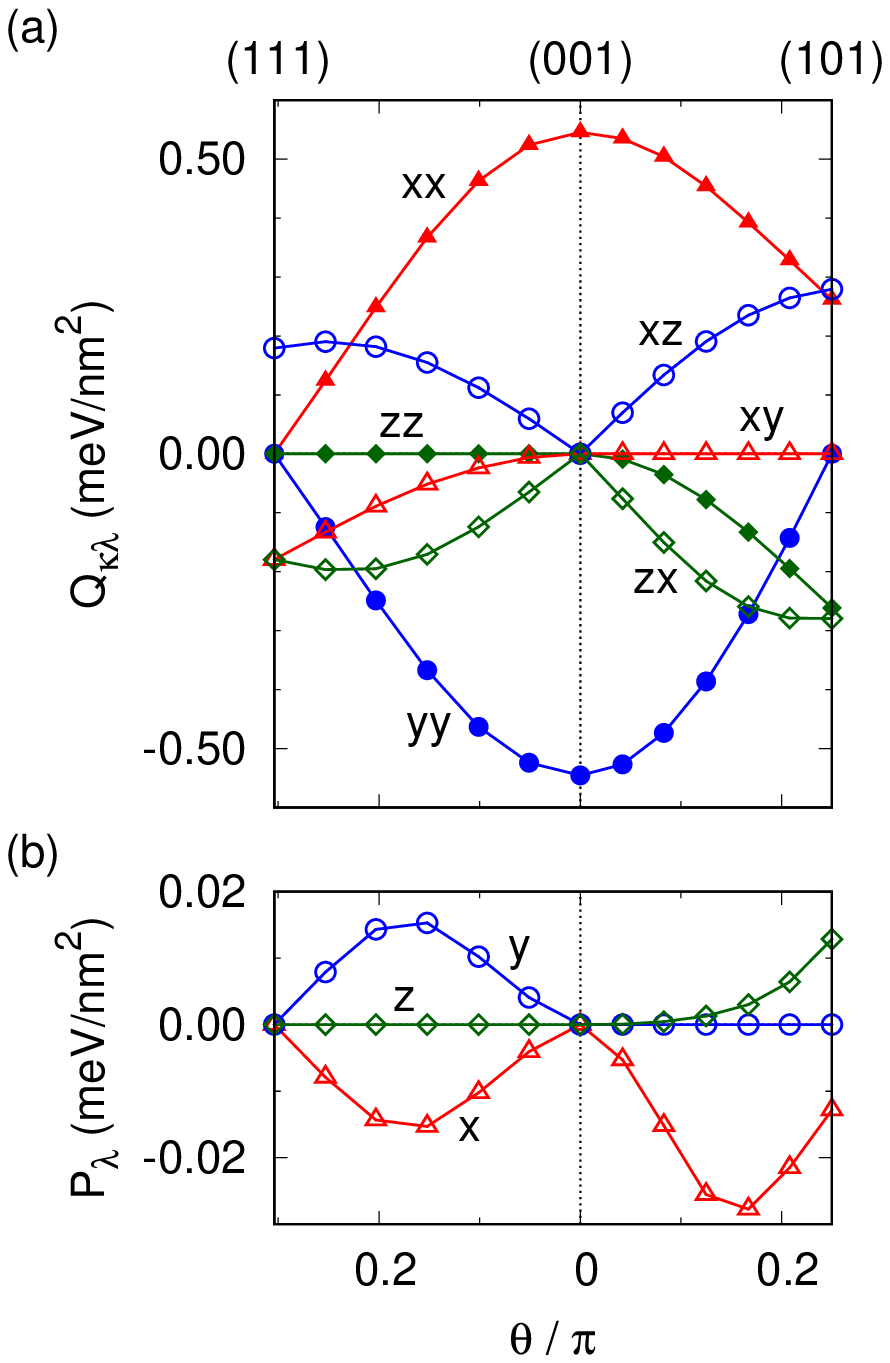}
\caption{Calculated spin currents in NiMnSb for magnetization
direction varying along the path (111) -- (001) -- (101):
components of the spin-current tensor $Q_{\kappa \lambda}$ (a)
and of the projection vector $P_\lambda$ (b).
For the missing components $Q_{\kappa \lambda}$, see text.
\label{fig_qnms}}
\end{center}
\end{figure}

More complete information about the spin currents can be
obtained by inspecting the spin current tensor 
$Q_{\kappa \lambda}$ and the projection vector $P_\lambda$
for magnetization direction $\mathbf{m}$ varying along the
path (111) -- (001) -- (101).
In terms of spherical angles $\theta$ and $\phi$, it holds
$m_x = \sin \theta \cos \phi$, $m_y = \sin \theta \sin \phi$, 
$m_z = \cos \theta$, and the segment (111) -- (001) is featured
by $\phi = \pi/4$, while the segment (001) -- (101) is featured
by $\phi = 0$.

\begin{figure}[h]
\begin{center}
\includegraphics[width=0.6\textwidth]{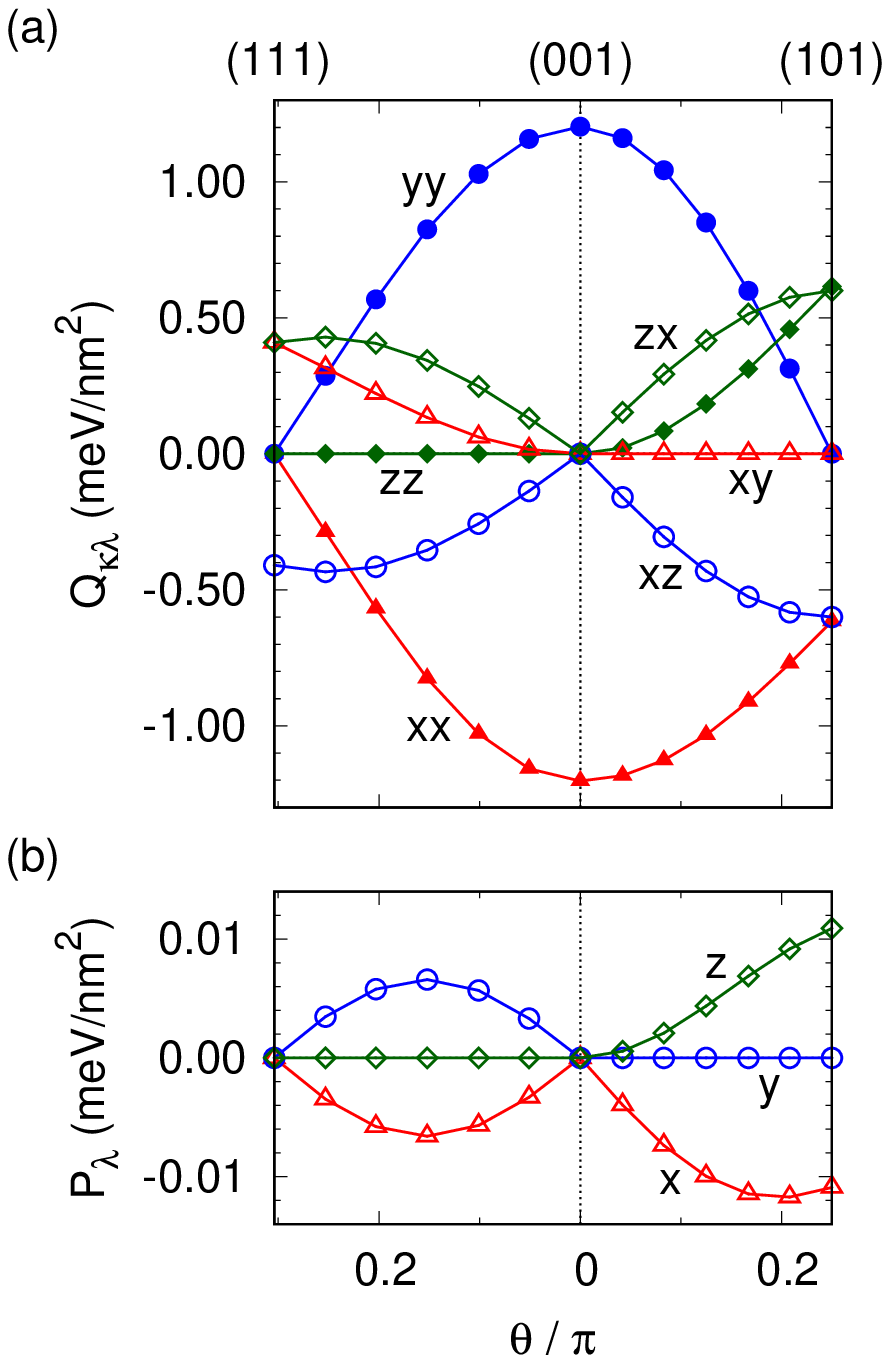}
\caption{The same as Fig.~\ref{fig_qnms}, but for CoMnFeSi.
\label{fig_qcmfs}}
\end{center}
\end{figure}

The resulting dependences are displayed in Fig.~\ref{fig_qnms}
and Fig.~\ref{fig_qcmfs} for NiMnSb and CoMnFeSi, respectively.  
The magnetic point group inside the segment (111) -- (001) is
$m'$ and the spin-current tensor has four independent nonzero
components ($Q_{xx} = -Q_{yy}$, $Q_{xy} = -Q_{yx}$, $Q_{xz} =
-Q_{yz}$, $Q_{zx} = -Q_{zy}$).
The magnetic point group inside the segment (001) -- (101) is
$2'$ leading to five components ($Q_{xx}$, $Q_{yy}$, $Q_{zz}$,
$Q_{xz}$, $Q_{zx}$).
All numerical results are consistent with the tensor shapes
obtained from the group theory (Section~\ref{ss_gta}).

It can also be seen that magnitudes of all components of the
projection vector $P_\lambda$ are substantially smaller than
the range of the spin-current values $Q_{\kappa \lambda}$.
This agrees fully with conclusions of Section~\ref{ss_psc} about
an approximate orthogonality between the spin polarization of the
intrinsic currents and the direction of magnetization.

\subsection{Dependence on magnetization direction\label{ss_dmd}}

Undoubtedly, deep understanding to the spin currents in
ferromagnets calls for an assessment of the dependence of the
tensor $Q_{\kappa \lambda}$ on the magnetization direction
$\mathbf{m}$.
Since the spin-current operator $\sigma_\kappa V_\lambda$ does
not change sign due to time reversal, this dependence has to
satisfy exactly the rule $Q_{\kappa \lambda} ( - \mathbf{m} ) =
Q_{\kappa \lambda} ( \mathbf{m} )$.
An approximate form of this dependence can be derived under
several assumptions.
We assume, e.g., that the Fermi energy of the system and the
magnitude of the exchange splitting in the Hamiltonian $H$,
Eq.~(\ref{eq_hdef}), do not change with varying $\mathbf{m}$.
If we confine ourselves to the simplest dependence compatible
with the mentioned rule, we have to consider a quadratic law
\begin{equation}
\tilde{Q}_{\kappa \lambda} ( \mathbf{m} ) = 
\sum_{\mu \nu} T_{\kappa \lambda \mu \nu} m_\mu m_\nu ,
\label{eq_tqdef}
\end{equation}
where the $T_{\kappa \lambda \mu \nu}$ are components of a
tensor of rank four.
This tensor is symmetric in the indices $\mu$ and $\nu$
($T_{\kappa \lambda \mu \nu} = T_{\kappa \lambda \nu \mu}$)
and it has to be invariant with respect to all elements of
the crystallographic point group of the underlying nonmagnetic
system.
This yields conditions
\begin{equation}
T_{\kappa \lambda \mu \nu} = 
| \alpha | \sum_{\kappa' \lambda' \mu' \nu'}
\alpha_{\kappa \kappa'} \alpha_{\lambda \lambda'}
\alpha_{\mu \mu'} \alpha_{\nu \nu'}
T_{\kappa' \lambda' \mu' \nu'} ,
\label{eq_tinv}
\end{equation}
where $\{ \alpha_{\mu \nu} \}$ are real orthogonal $3 \times 3$
matrices representing the elements $\alpha$ of the point group
and $| \alpha |$ denotes determinant of $\alpha$.
These conditions and the symmetry property define the shape
of the tensor $T_{\kappa \lambda \mu \nu}$.
The numbers of independent components of this tensor for all
noncentrosymmetric point groups are listed in
Table~\ref{tab_tr4}.

\begin{table}
\caption{%
Numbers of independent components of the tensor
$T_{\kappa \lambda \mu \nu}$ for all noncentrosymmetric
crystallographic point groups.
\label{tab_tr4}}
\begin{center}
\begin{tabular}{lc}
\hline
point group \phantom{m} & no.\ of components \\
\hline
$1$ & 54 \\
$2$ & 28 \\
$m$ & 26 \\
$3$ & 18 \\
$222$ & 15 \\
$4$, $\bar{4}$ & 14 \\
$2mm$ & 13 \\
$6$ & 12 \\
$32$ & 10 \\
$3m$, $422$ & 8 \\
$\bar{4}2m$, $622$ & 7 \\
$4mm$, $\bar{6}$ & 6 \\
$6mm$, $23$ & 5 \\
$\bar{6}m2$, $432$ & 3 \\
$\bar{4}3m$ & 2 \\
\hline
\end{tabular}
\end{center}
\end{table}
 
One can find that in the case of the point group $\bar{4}3m$
(full symmetry group of a regular tetrahedron), there are
only two independent components of the tensor  
$T_{\kappa \lambda \mu \nu}$; the corresponding dependence,
Eq.~(\ref{eq_tqdef}), reduces to
\begin{eqnarray}
\tilde{Q}_{xx} & = & A (m^2_y - m^2_z ) ,
\nonumber\\
\tilde{Q}_{yy} & = & A (m^2_z - m^2_x ) ,
\nonumber\\
\tilde{Q}_{zz} & = & A (m^2_x - m^2_y ) ,
\nonumber\\
\tilde{Q}_{xy} & = & - \tilde{Q}_{yx} \ = \ A' m_x m_y ,
\nonumber\\
\tilde{Q}_{yz} & = & - \tilde{Q}_{zy} \ = \ A' m_y m_z ,
\nonumber\\
\tilde{Q}_{zx} & = & - \tilde{Q}_{xz} \ = \ A' m_x m_z ,
\label{eq_tqres}
\end{eqnarray}
where $A$ and $A'$ are two constants. 
Note that the constants $A$ and $A'$ define separately the
diagonal and nondiagonal components of the tensor
$\tilde{Q}_{\kappa \lambda}$.

\begin{figure}[h]
\begin{center}
\includegraphics[width=0.45\textwidth]{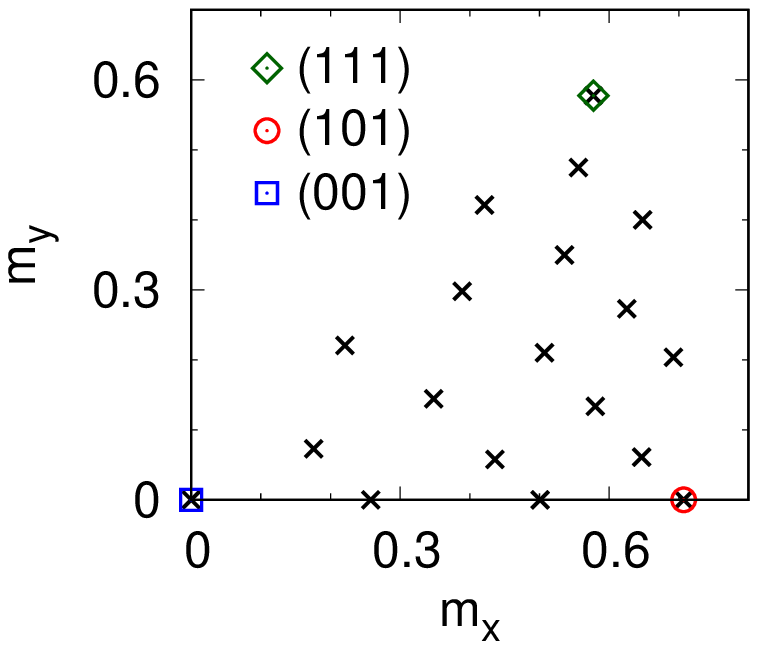}
\caption{Projection of the sampling directions $\mathbf{m}$
on the $x - y$ plane.
\label{fig_magdir}}
\end{center}
\end{figure}

In order to check the validity of the approximate dependence,
Eq.~(\ref{eq_tqres}), calculations for a number of unit
vectors $\mathbf{m}$ were carried out and the resulting values
of $Q_{\kappa \lambda} ( \mathbf{m} )$ were compared to
the values of $\tilde{Q}_{\kappa \lambda} ( \mathbf{m} )$ with
the constants $A$ and $A'$ obtained from a least-squares fit.
The sampling vectors $\mathbf{m}$ are shown in
Fig.~\ref{fig_magdir}; they were chosen to scan the region
among the three high-symmetry directions of the system.

The resulting constants are: $A = -0.537$ meV/nm$^2$ and
$A' = -0.554$ meV/nm$^2$ for NiMnSb while $A = 1.210$ meV/nm$^2$
and $A' = 1.212$ meV/nm$^2$ for CoMnFeSi.
One can see that $A \approx A'$ for both systems.
It can also be observed that for $A = A'$, the approximate
dependence, Eq.~(\ref{eq_tqres}), satisfies a sum rule
$\sum_\kappa m_\kappa \tilde{Q}_{\kappa \lambda} ( \mathbf{m} )
= 0$ for all $\lambda$ and $\mathbf{m}$, so that the
magnetization direction is exactly orthogonal to the spin
polarization of the approximate spin currents.
In other words, the small relative difference between $A$ and
$A'$ reflects the small values of the projection vector
$P_\lambda$ (Section~\ref{ss_scha}).

The accuracy of the developed scheme can be quantified, e.g.,
by comparing the calculated and fitted spin currents for the
high-symmetry directions, which indicates a very good agreement,
see Table~\ref{tab_qcalc}. 
A more systematic comparison should include all sampling
vectors $\mathbf{m}$ used for the fit (Fig.~\ref{fig_magdir}).
A relative deviation can be defined as the maximum over all
$\kappa$, $\lambda$, and $\mathbf{m}$ of the quantity
$| Q_{\kappa \lambda} ( \mathbf{m} ) - \tilde{Q}_{\kappa \lambda}
( \mathbf{m} )| / | A |$.
This deviation comes out 0.037 for NiMnSb and 0.0094 for
CoMnFeSi, so that it does not exceed a few percent in the
studied systems.
This analysis proves that the derived quadratic formula with
two constants, Eq.~(\ref{eq_tqres}), provides a good starting
point for more accurate approximations of the full dependence
of the spin-current tensor on magnetization direction.

%\newpage

\section{Conclusions\label{s_conc}}

This pilot theoretical study has been devoted to spin currents
which are not induced by external electric fields but arise only
due to spin-orbit interaction in the absence of space-inversion
symmetry.
The group analysis revealed that these spin currents appear
nearly in all bulk systems with noncentrosymmetric magnetic point
groups compatible with ferromagnetism; the only exceptions
are systems with point groups $\bar{6}$ and $\bar{6}m'2'$,
where the symmetry also leads to vanishing spin-orbit torques
due to external electric fields.

The study has focused on systems with weak spin-orbit
interaction as compared to exchange splitting.
In this regime, the analysis of a perturbation expansion proved
that spin polarization of the intrinsic currents is nearly
perpendicular to the direction of magnetization.
This property differs from that of currents generated by external
electric fields, where the spin polarization of the current
contains a substantial component parallel to the magnetization.

The general conclusions drawn have been corroborated by
\emph{ab initio} calculations for NiMnSb and CoMnFeSi compounds,
for which the dependence of the spin-current tensor on the
magnetization direction has also been studied.
A simple quadratic approximation of this dependence has been
derived and its accuracy checked; relative deviations not
exceeding a few percent have been found for both systems.
In view of the close relation between the intrinsic spin currents
and the Dzyaloshinskii-Moriya interaction, one can expect that
results of similar studies for other ferromagnets will serve
as an input for micromagnetic simulations of systems that
are collinear on a short length scale, but noncollinear on
a longer scale, such as spin spirals, magnetic domain walls,
and skyrmions. 

\subsection*{Acknowledgments}

This work was supported financially by the Czech Science
Foundation (grant No.\ 23-04746S).
The author thanks Dr.\ Karel Carva and Dr.\ Alberto
Marmodoro for stimulating discussions.

%\newpage

\appendix

\section{One-dimensional tight-binding model\label{app_1d}}

\begin{figure}[h]
\begin{center}
\includegraphics[width=0.4\textwidth]{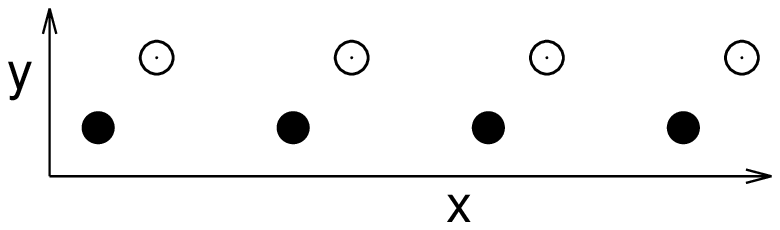}
\caption{Structure of the one-dimensional chain with magnetic
(full circles) and nonmagnetic (open circles) sites.
\label{fig_chain}}
\end{center}
\end{figure}

The structure of the one-dimensional model is depicted in
Fig.~\ref{fig_chain}: an infinite periodic chain along the
$x$ axis comprising one magnetic site and one nonmagnetic
site in the unit cell, with all sites lying in the $x - y$
plane.

The tight-binding Hamiltonian assumes three $p$-type orbitals
attached to each magnetic site and one $s$-type orbital
attached to each nonmagnetic site.
The exchange splitting at each magnetic site has a form
$\mathbf{b} \cdot \bm{\sigma}$, where $\mathbf{b} = 
( b_x , b_y , b_z )$ is the exchange field, while SO
interaction at the same site has a form 
$\xi \mathbf{L} \cdot \bm{\sigma}$, where $\mathbf{L} = 
( L_x , L_y , L_z )$ is the operator of orbital momentum and
$\xi$ denotes the strength of SO interaction.
Nonzero hopping integrals are assumed between:
(i) the $p$ orbitals of the nearest-neighbor magnetic sites,
(ii) each $s$ orbital and its two nearest $p_x$ orbitals, and
(iii) each $s$ orbital and its first nearest $p_y$ orbital.
After a downfolding of the $s$ orbitals, an effective Hamiltonian
for the magnetic sites with $p$ orbitals is obtained.
Its lattice Fourier transformation can be represented by a
matrix $H_{\mu s , \mu' s'} (k)$, where $k$ is the
one-dimensional reciprocal-space variable, $\mu, \mu' \in
\{ x , y , z \}$ refer to the particular $p$ orbital, and
$s, s' \in \{ \uparrow , \downarrow \}$ are the spin indices.

\begin{figure}[h]
\begin{center}
\includegraphics[width=0.35\textwidth]{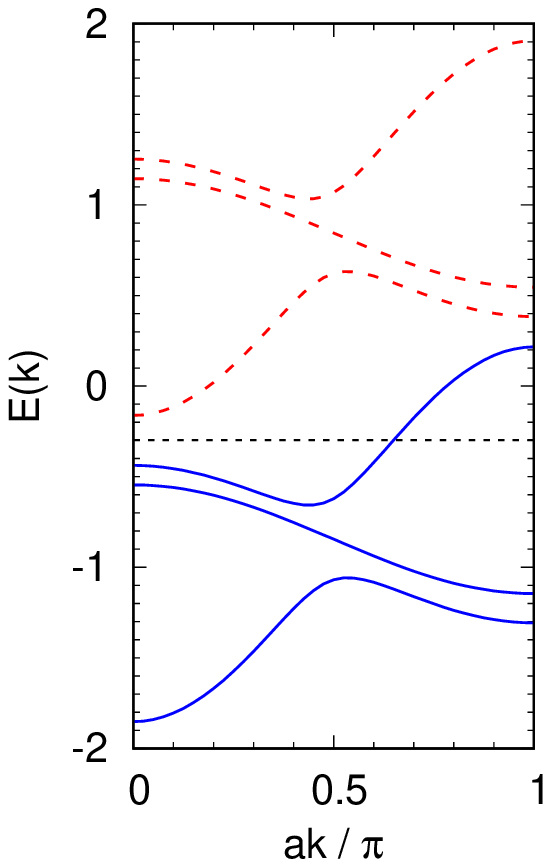}
\caption{Bandstructure of the one-dimensional model with
null SO interaction ($\xi = 0$): majority bands (full lines)
and minority bands (dashed lines).
The horizontal short-dashed line denotes the position of the
Fermi level $E_\mathrm{F}$.
\label{fig_bands}}
\end{center}
\end{figure}

The resulting Hamiltonian has a form of Eq.~(\ref{eq_hdef}),
where both local terms lead to $k$-independent matrices
\begin{equation}
H^\mathrm{ex}_{\mu s , \mu' s'} = \delta_{\mu \mu'}
\sum_\lambda b_\lambda (\sigma_\lambda)_{s s'}  
\label{eq_1dhex}
\end{equation}
and 
\begin{equation}
H^\mathrm{SO}_{\mu s , \mu' s'} = \xi \sum_\lambda
(L_\lambda)_{\mu \mu'} (\sigma_\lambda)_{s s'} ,
\label{eq_1dhso}
\end{equation}
where we set $(L_\lambda)_{\mu \mu'} = -i
\varepsilon_{\mu \mu' \lambda}$ with
$\varepsilon_{\mu \mu' \lambda}$ denoting the Levi-Civita tensor.
The spin-independent term $\bar{H}$ becomes a $k$-dependent
matrix,
\begin{eqnarray}
\bar{H}_{\mu s , \mu' s'} (k) & = & \delta_{s s'} \,
 W_{\mu \mu'} (k) ,
\nonumber\\
W_{xx} (k) & = &  -2 h_{xx} \cos(ak) ,
\nonumber\\
W_{yy} (k) & = &  2 h_{yy} \cos(ak) ,
\nonumber\\
W_{zz} (k) & = &  2 h_{zz} \cos(ak) ,
\nonumber\\
W_{yx} (k) & = &  h_n - h'_n \exp(iak) ,
\nonumber\\
W_{zx} (k) & = &  W_{yz} (k) \ = \ 0 ,
\label{eq_1dhbar}
\end{eqnarray}
where all omitted matrix elements can be restored from
$W_{\mu' \mu} (k) = W^\ast_{\mu \mu'} (k)$. 
Here $a$ is the lattice parameter and $h_{xx}$, $h_{yy}$,
$h_{zz}$, $h_n$, and $h'_n$ are hopping parameters of the model.
The representation of the velocity operator $V_x$ is obtained
from the relation $(V_x)_{\mu s , \mu' s'} (k) = \delta_{s s'}
\, dW_{\mu \mu'} (k) / dk$. 

The calculations were carried out with following values of the
model parameters: $b_x = 0.195$, $b_y = 0.26$, $b_z = 0.78$, 
$h_{xx} = 0.5$, $h_{yy} = 0.2$, $h_{zz} = 0.15$, $h_n = 0.2$,
and $h'_n = 0.1$, and with the Fermi energy
$E_\mathrm{F} = - 0.3$; the bandstructure of the model for
$\xi = 0$ is shown in Fig.~\ref{fig_bands}.
For evaluation of the spin currents, Eq.~(\ref{eq_qdef}),
a Lorentzian broadening of the eigenvalues with a parameter
$\gamma = 0.01$ was used and the strength of SO interaction was
varied in the interval $-0.05 \leq \xi \leq 0.05$, see
Fig.~\ref{fig_scaling}.

%\newpage

%\bibliography{bb_}

\providecommand{\noopsort}[1]{}\providecommand{\singleletter}[1]{#1}%

\end{document}